\documentclass[useAMS,usenatbib]{mn2e}
\usepackage{graphicx}
\usepackage{txfonts}
\usepackage[dvipsnames]{color}
\usepackage{ulem}
\def\aap{A\&A}
\def\apj{ApJ}
\def\apjl{ApJL}
\def\apjs{ApJS}
\def\mnras{MNRAS}

\def\pasj{PASJ}
\def\nat{Nature}

\usepackage{soul}
\usepackage[dvipsnames]{xcolor}

\title[Linear Theory of the RTI in a Relativistic Flow]{Linear Theory of the Rayleigh--Taylor Instability at a Discontinuous Surface of a Relativistic Flow}
\author[J. Matsumoto, M. A. Aloy and M. Perucho]{Jin Matsumoto$^{1}$\thanks{E-mail:jin.matsumoto@riken.jp} ,
Miguel A. Aloy$^{2}$ and Manel Perucho$^{2,3}$ \\
$^{1}$Astrophysical Big Bang Laboratory, Riken, Wako, 351-0198, Japan\\
$^{2}$Departament d'Astronomia i Astrof\'{\i}sica, Universitat de Val\`encia, C/ Dr. Moliner, 50, 46100, Burjassot, Val\`encia, Spain.\\
$^{3}$Observatori Astron\`omic, Universitat de Val\`encia, C/ Catedr\`atic Jos\'e Beltr\'an 2, 46980, Paterna, Val\`encia, Spain.}
\begin{document}
\date{Accepted xxx, Received yyy, in original form zzz}
\pagerange{\pageref{firstpage}--\pageref{lastpage}} \pubyear{2017}
\maketitle
\label{firstpage}
\begin{abstract}
We address the linear stability of a discontinuous surface of a relativistic 
flow in the context of a jet that oscillates radially as it propagates. The 
restoring force of the oscillation is expected to drive a Rayleigh--Taylor 
instability (RTI) at the interface between the jet and its cocoon. We perform 
a linear analysis and numerical simulations of the growth of the RTI in the 
transverse plane to the jet flow with a uniform acceleration. In this system, 
an inertia force due to the uniform acceleration acts as the restoring force 
for the oscillation. We find that not only the difference in the inertia between 
the two fluids separated by the interface but also the pressure at the interface 
helps to drive the RTI because of a difference in the Lorenz factor across the 
discontinuous surface of the jet. The dispersion relation indicates that the 
linear growth rate of each mode becomes maximum when the Lorentz factor 
of the jet is much larger than that of the cocoon and the pressure at the jet 
interface is relativistic. By comparing the linear growth rates of the RTI in the 
analytical model and the numerical simulations, the validity of our analytically 
derived dispersion relation for the relativistic RTI is confirmed.
\end{abstract}

\begin{keywords}
galaxies: jets --- instabilities --- methods: analytical --- methods: numerical --- relativistic processes
\end{keywords}

\section{Introduction}
The interface between two fluids of different densities is unstable if
the heavier fluid is supported above the lighter one against gravity,
or equivalently if the lighter fluid accelerates the heavier one. This
instability is known as the Rayleigh--Taylor instability
\citep[RTI,][]{Rayleigh1900, Taylor50}.  It is a fundamental process
in hydrodynamics and plays an important role in many astrophysical 
contexts. 

In massive stars, the composition interfaces between the hydrogen- and
helium-rich layers and between the helium-rich layer and C+O core are
Rayleigh--Taylor (RT) unstable after a supernova shock wave passes
through them \citep{Chevalier76, Ebisuzaki89, Hachisu92, Ono13, Mao15}. 
The growth of such RTIs is thought to be a promising mechanism for the
material mixing or the penetration of lighter and heavier elements
into the neighboring layers in supernova explosions
\citep{Kifonidis03, Wongwathanarat15}. 

In addition to the above, the possibility of RTIs in supernova remnants 
has been discussed. The ejected stellar envelope associated with a 
supernova explosion is decelerated by the swept-up interstellar medium. 
The interface separating the denser ejecta from the shocked interstellar 
medium is also RT unstable \citep{Ferrand12, Warren13, Obergaulinger14a}. 
The growth of RTIs is responsible for the finger-like structures found in 
supernovae remnants. The stretching of the magnetic field lines by these 
RT fingers may help to amplify the magnetic field, thereby explaining the 
observed synchrotron emission from the thin shell of the supernova 
remnant \citep{Jun95, Jun96, Guo12, Obergaulinger14b}. 

The magnetic RTI \citep{Hillier16} itself is also expected to be
important for the emergence of magnetic flux from the solar interior
\citep{Isobe05, Isobe06} and the buoyancy of bubbles in the solar
prominence \citep{Hillier11, Hillier12}. The magnetic tension
suppresses the short-wavelength modes along the magnetic field
except when the wavevector is perpendicular to the magnetic field. 
As a result of the different development of the magnetic
RTI along and across the magnetic field lines, there is a
preferential formation of astrophysical features elongated in the
direction of the magnetic field.

In a super-Eddington outflow from an accretion disk, in
which the outward acceleration due to the radiation force is larger
than the inward gravitational pull of the central black hole, the radiation 
force drives RTIs. These instabilities are located at the photosphere 
of the accretion disk, since the density decreases outwardly there 
\citep{Takeuchi13, Takeuchi14}. The radiation force acts as an external 
inertia force in the driving mechanism of this instability. The growth of
radiation-driven RTIs may be responsible for the formation of clumpy
structures in super-Eddington outflows.

The relativistic RTI is a key process in the dynamics of high-energy 
astrophysics. The contact discontinuity in a relativistic shell propagating 
through the interstellar medium is subjected to an RTI in the context of 
a gamma-ray burst \citep[GRB,][]{Duffell13, Duffell14}. The physical 
reason for the onset of such an RTI is similar to that of the supernova 
remnant, except for the velocity of the ejecta shell. This instability may 
be responsible for amplifying the magnetic field via small-scale turbulent 
dynamo facilitating the synchrotron emission for the GRB afterglow.

An unstable interface appears in the interaction between stellar and
relativistic pulsar winds \citep{Bosch-Ramon15,Christie16}. Since the 
low-density shocked pulsar wind accelerates the denser shocked stellar 
wind, RTIs occur at the interface between the shocked winds. The 
development of such RTIs has a large impact on the evolution of the 
shocked winds.

Besides the aforementioned topics, the growth of RTIs at the interface 
between a relativistic jet and its surrounding medium impacts on the 
stability of the jet structure when the jet either expands radially because 
of the centrifugal force \citep{Meliani07, Meliani09, Millas17} or oscillates 
radially because of a pressure gradient \citep{Matsumoto13, Toma17}. 
The stability of relativistic jets is important for the acceleration/deceleration 
and collimation mechanisms of the GRB, active galactic nucleus (AGN), 
and microquasar jets. 

The stability of the jet interface is also related to the inhomogeneity of the 
jet and to the evolution of turbulence inside/outside the jet. These processes 
affect the radiative output from the jet associated with particle and/or photon 
acceleration. Multiple outflow layers inside a relativistic jet are essential for 
reproducing the typical observed spectra of GRBs \citep{Ito14}. The 
development of turbulence inside the jet is an important issue in any 
discussion of the mechanism for efficient particle acceleration in the context 
of GRBs \citep{Asano15b} and blazars \citep{Asano15a, Inoue16}.

Since there is a velocity shear at the interface between the jet and
external medium, a promising mechanisms for destabilizing the jet 
interface are the Kelvin--Helmholtz instability (KHI) and the shear 
driven instability (e.g., \citealt{Urpin02, Aloy02}).  Many authors have 
investigated the growth of KHIs at the relativistic flow interface both
analytically and numerically \citep[e.g.][]{Turland76, Blandford76,
Ferrari78, Hardee79, Hardee98, Hardee01, Perucho04, Perucho05,
Perucho07, Mizuno07, Rossi08, Perucho10}. By comparison, the growth
of RTIs at such an interface is not still well understood. Even the
general conditions for the onset of the RTI at the jet interface are
unclear at present.

The dispersion relation for the relativistic RTI was derived by \citet{Allen84} 
and \citet{Duffell11}. However, those studies were limited to the non-relativistic 
flow of relativistically hot gas. \citet{Meliani09} derived the condition for RTI 
onset at the interface of the relativistic jet by using an approximate dispersion 
relation. 

\citet{Levinson10} performed a stability analysis of the two-shock solution 
\citep{Nakamura06} describing the interaction of relativistic ejecta with an 
ambient medium and showed that the contact discontinuity between the 
shocked ejecta and the shocked ambient medium was RT unstable. However, 
their analysis is not directly applicable to the interface of a relativistic jet. This 
is because the direction of the normal vector of the contact discontinuity is 
perpendicular to the relativistic flow in the jet--external-medium system 
(considered as a slice transversal to the jet flow, see Fig.~1b) whereas it is 
parallel to the relativistic flow in the interaction between the relativistic ejecta 
and the ambient medium. 

In this paper, we study the general conditions for the onset and
growth of the relativistic RTI at the discontinuous surface of the
relativistic flow. For this purpose, we perform a linear analysis and
numerical simulations of the RTI in a simple jet--cocoon-medium
system, and we compare the analytically derived growth rate to that
estimated numerically. Our findings, such as the dispersion relation
and the growth rate of the relativistic RTI, are also applicable to
analyzing the stability of the interface between stellar and relativistic 
pulsar winds \citep{Bosch-Ramon15,Christie16} as well as the stability 
prospects of the formation of anomalous shear layers in relativistic 
jets (see, e.g., \citealt{AR06, AM08, Mizuno08, Zenitani10}).

This paper is organized as follows. In Section~2, we present the linear 
analysis and derive the dispersion relation of the relativistic RTI. In 
Section~3, we compare RTI growth rates between the analytical model 
and the numerical simulations. Finally, we summarize and discuss our 
findings in Section~4.

\begin{figure*}
\begin{center}
\scalebox{0.349}{{\includegraphics{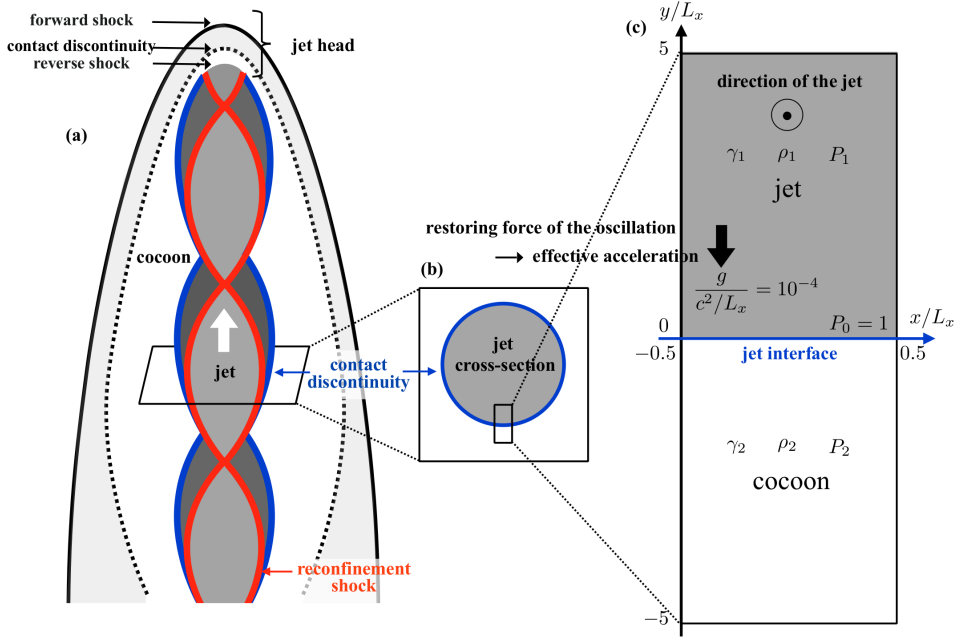}}} 
\caption{Panel~(a): Schematic picture of the jet propagation. Panel~(b): 
Cross-section of the jet. Panel~(c): Geometry of the jet--cocoon medium 
system in the linear stability analysis and numerical simulations. Here, $L_x$, 
$P_0$ and $c$ are the length of the computational domain in the $x-$direction, 
$P_0$ the pressure normalization and $c$ the speed of light in vacuum.
}
\label{fig1}
\end{center}
\end{figure*}

\section{Linear analysis}
\subsection{Physical assumptions and basic equations}
We investigate the stability of the radially oscillating interface between 
the jet and the cocoon medium. This radial motion of the jet is excited 
naturally by the pressure mismatch between the jet and its surrounding 
medium (i.e. the cocoon in typical AGN jets) when the jet propagates 
through an ambient medium \citep{Sanders83, Matsumoto12}. This 
oscillatory motion of the jet is the origin of the formation of the 
reconfinement shocks inside the jet. In the rest frame of the decelerating 
jet interface that is expanding radially (i.e., in the direction perpendicular 
to the jet axis), an inertia force acts on the interface and is directed 
outwards. Therefore, the jet flow is driven against the cocoon in the 
direction opposite to the inertia force in this frame. We point out that the 
direction of the jet interface is perpendicular to the relativistic flow of 
the jet. Figure~\ref{fig1} shows a schematic picture of the jet--cocoon
system that we consider in this study. Next, we derive the dispersion
relation of the RTI in this system following the standard procedure
for the RTI in classical hydrodynamics \citep{Chandrasekhar61}.

Assuming that magnetic fields are dynamically negligible, the jet--cocoon
system can be modeled as an ideal gas subject to the equations of
relativistic hydrodynamics, which can be suitably written as
\begin{eqnarray}
\frac{\partial}{\partial t}(\gamma \rho) + \nabla \cdot (\gamma \rho \mbox{\boldmath $v$}) = 0 \;, \label{eq: mass conservation}
\end{eqnarray}
\begin{eqnarray}
\gamma^2 \rho h \biggl [ \frac{\partial \mbox{\boldmath $v$}}{\partial t} + (\mbox{\boldmath $v$} \cdot \nabla) \mbox{\boldmath $v$} \biggr ] 
= -\nabla P - \frac{\mbox{\boldmath $v$}}{c^2} \frac{\partial P}{\partial t} + 
\gamma^2 \rho h \mbox{\boldmath $g$} \;, \label{eq: equation of motion1}
\end{eqnarray}
\begin{eqnarray}
\frac{{\rm d} s}{{\rm d}t} = 0 \;, \label{eq: adiabatic condition1}
\end{eqnarray}
where
\begin{eqnarray}
h := 1 + \frac{\Gamma}{\Gamma -1} \frac{P}{\rho c^2} \label{eq: specific enthalpy}
\end{eqnarray}
and
\begin{eqnarray}
s := {\rm log} \biggl ( \frac{P^{1/\Gamma -1}}{\rho^{\Gamma/\Gamma -1}} \biggr ) = \frac{1}{\Gamma -1} {\rm log} \frac{P}{\rho^\Gamma} \;.
\end{eqnarray}
These are the continuity equation (\ref{eq: mass conservation}), the
equation of motion (\ref{eq: equation of motion1}) and the entropy
conservation equation (\ref{eq: adiabatic condition1}). Here, $\rho$
is the rest-mass density, $P$ is the pressure, $h$ is the specific
enthalpy, $s$ is the entropy density, $\Gamma$ is the ratio
of specific heats, $\mbox{\boldmath $v$}$ is the velocity vector,
$\gamma := (1-v^2/c^2)^{-1/2}$ is the Lorentz factor, $c$ is the
speed of light and $\mbox{\boldmath $g$}$ is the acceleration vector
for the inertia force.

In order to simplify the analysis, we assume that the jet radius is
large enough that the jet interface can locally be regarded as a
planar, rather than a quasi-cylindric surface. With this simplification, 
we neglect the impact of the curvature of the jet interface on the growth 
of the RTI.  Moreover, this approximation allows us to obtain the sought 
dispersion relation employing Cartesian coordinates $(x,y,z)$. The 
assumed geometry of the jet--cocoon system in the forthcoming linear 
stability analysis and numerical simulations is sketched in Fig.~\ref{fig1}c.  
The jet beam is shaped by the cocoon when a uniform acceleration is 
directed in the negative $y-$direction. The unperturbed jet beam moves 
in the positive $z-$axis. The interface between the jet and the cocoon is 
initially located at $y=0$. We consider the dynamics only in the $x$--$y$ 
plane by assuming that derivatives of the physical variables in the 
$z-$direction are zero although the $z-$component of velocity itself is 
taken into account.  Neglecting the variations in the $z-$direction means 
assuming that both, the temporal and the spatial variations of the physical 
variables along the $z-$direction are much smaller than in the $x$--$y$ 
plane. We will check the validity of this assumption in light of the 
developments of Sect.~\ref{sec:dispersion}. We anticipate that the KHI 
can not grow under the previous assumption. In addition, we neglect the 
temporal variation of the pressure in equation~(\ref{eq: equation of motion1}) 
because the pressure gradient force is dominant in the process of the jet 
oscillation and counterbalances the inertia force in the rest frame of the 
jet interface. The $z-$component of velocity, $v_z$, is not perturbed under 
these assumptions. This is because there is no external force in the 
$z-$direction: ${\rm d} v_{z}/{\rm d}t = 0$. The $z-$component of velocity 
contributes only to the Lorentz factor.  Since the motion of the fluid in the 
$x$--$y$ plane is much slower than the speed of sound, we assume the 
fluid is incompressible. Writing the incompressibility condition in 
component form and added to the governing equations 
(\ref{eq: mass conservation})--(\ref{eq: adiabatic condition1}) we obtain:
\begin{eqnarray}
\frac{\partial}{\partial t}(\gamma \rho) + v_{x}\frac{\partial}{\partial x}(\gamma \rho) + 
v_{y}\frac{\partial}{\partial y}(\gamma \rho) = 0 \;, \label{eq: incompressible condition1} 
\end{eqnarray}
\begin{eqnarray}
\frac{\partial v_{x}}{\partial x} + \frac{\partial v_{y}}{\partial y} = 0 \;, \label{eq: continuity equation}
\end{eqnarray}
\begin{eqnarray}
\gamma^2 \rho h \biggl ( \frac{\partial v_{x}}{\partial t} + v_{x}\frac{\partial v_{x}}{\partial x} + 
v_{y}\frac{\partial v_{x}}{\partial y} \biggr ) = -\frac{\partial P}{\partial x} \;, 
\end{eqnarray}
\begin{eqnarray}
\gamma^2 \rho h \biggl ( \frac{\partial v_{y}}{\partial t} + v_{x}\frac{\partial v_{y}}{\partial x} + 
v_{y}\frac{\partial v_{y}}{\partial y} \biggr ) = -\frac{\partial P}{\partial y} - \gamma^2 \rho h g \;, 
\end{eqnarray}
\begin{eqnarray}
\frac{\partial s}{\partial t} + v_{x}\frac{\partial s}{\partial x} + v_{y}\frac{\partial s}{\partial y} = 0 \;.
\label{eq: adiabatic condition2}
\end{eqnarray}
Equation~(\ref{eq: incompressible condition1}) is the incompressibility 
condition for the relativistic gas restricted to motion in the $x$--$y$ 
plane. 

\subsection{Equilibrium state for the linear stability analysis}
We assume that the jet--cocoon-medium system is initially in 
hydrostatic equilibrium in the $x$--$y$ plane. The pressure 
gradient counterbalances the inertia force in the $y-$direction;
\begin{eqnarray}
\frac{\partial P}{\partial y} = - \gamma^2 \rho h g \;. \label{eq: balancing equation}
\end{eqnarray}
Assuming that $\rho$ and $v_z$ are uniform and that in the unperturbed
estate $v_x=v_y=0$ in both the jet and in the cocoon regions, the
pressure distribution in the initial equilibrium is obtained as
follows:
\begin{eqnarray}
P = P_0 e^{-y/H} + \frac{\Gamma -1 }{\Gamma} \rho c^2 (e^{-y/H} - 1) \;, \label{eq: pressure distribution}
\end{eqnarray}
where $P_0$ is the pressure at $y=0$ and 
\begin{eqnarray}
H := \frac{\Gamma - 1}{\Gamma} \frac{c^2}{\gamma^2 g} \; \label{eq: pressure scale height1}
\end{eqnarray}
is the pressure scale height determined by the acceleration 
and the Lorentz factor. Since the inertia force has its origin 
in the radially oscillating motion of the jet, assuming the 
amplitude of the jet oscillation is roughly equal to the jet radius, 
the magnitude of the acceleration $g$ is estimated as follows:
\begin{eqnarray}
g \sim \frac{r_{\rm jet}}{\tau_{\rm osci}^2} \;. \label{eq: effective gravity}
\end{eqnarray}
Here $\tau_{\rm osci}$ is the typical oscillation time of the jet and is given by sound
crossing time of the jet radius \citep{Matsumoto12}:
\begin{eqnarray}
\tau_{\rm osci} = \gamma_{\rm jet} r_{\rm jet}/C_{\rm s} \;. \label{eq: oscillation time}
\end{eqnarray}
Note that $\gamma_{\rm jet}$ is the typical Lorentz factor of the jet
and must be included to compute the sound crossing time in the 
laboratory frame. The sound speed $C_{\rm s}$ is maximal and
equal to $c/\sqrt{3}$ if the jet is relativistically hot. 
From equations~(\ref{eq: pressure scale height1})--(\ref{eq: oscillation
time}), we estimate the pressure scale height in the jet region to
be larger than the jet radius:
\begin{eqnarray}
H \gtrsim r_{\rm jet} \;. \label{eq: pressure scale height2}
\end{eqnarray}
Since we neglect the impact of the curvature of the jet radius in this
study, everything located at distances $y\ge H$ can be considered 
as ``far'' from the jet interface.

\subsection{Linearized equations}
We investigate the stability of the initial hydrostatic
equilibrium between the jet
and the cocoon by disturbing the system slightly and following its
evolution. We consider the actual density, pressure and velocity
components in the perturbed state to be $\rho + \delta \rho$,
$P + \delta P$, $\delta v_{x}$ and $\delta v_{y}$, respectively. Note
that since the perturbed velocity is perpendicular to the unperturbed
velocity that has only a $z-$component, the linearized Lorentz factor
in the perturbed state corresponds to that in the unperturbed
state. When equations~(\ref{eq: incompressible condition1})--(\ref{eq: 
adiabatic condition2}) are linearized, they become
\begin{eqnarray}
\gamma \frac{\partial \delta \rho}{\partial t} + \delta v_{y}\frac{\partial (\gamma \rho)}{\partial y} = 0 \;,
\label{eq: incompressible condition2}
\end{eqnarray}
\begin{eqnarray}
\frac{\partial \delta v_{x}}{\partial x} + \frac{\partial \delta v_{y}}{\partial y} = 0 \;,
\end{eqnarray}
\begin{eqnarray}
\gamma^2 \rho h \frac{\partial \delta v_{x}}{\partial t} = - \frac{\partial \delta P}{\partial x} \;,
\end{eqnarray}
\begin{eqnarray}
\gamma^2 \rho h \frac{\partial \delta v_{y}}{\partial t} = - \frac{\partial \delta P}{\partial y} 
- \gamma^2 \biggl ( \delta \rho + \frac{\Gamma}{\Gamma -1} \frac{\delta P}{c^2} \biggr ) g \;,
 \label{eq: equation of motion2}
\label{eq: }
\end{eqnarray}
\begin{eqnarray}
\frac{1}{P} \biggl ( \frac{\partial \delta P}{\partial t} + \delta v_{y} \frac{\partial P}{\partial y} \biggr )
- \frac{\Gamma}{\rho} \biggl ( \frac{\partial \delta \rho}{\partial t} + \delta v_{y} \frac{\partial \rho}{\partial y} \biggr ) = 0 \;.
\label{eq: adiabatic condition3}
\end{eqnarray}
As discussed by \citet{Allen84}, this set of equations over-constrains
the problem in classical cases \citep{Chandrasekhar61, Allen84}
because of the additional assumption of fluid incompressibility. Therefore, 
in those studies, the energy equation was not taken into account because 
it contains no extra information. However, the situation here is slightly 
different from that in previous work. In addition to a density jump, there is 
a difference in the Lorentz factor across the interface between two fluids. 
Using equation (\ref{eq: incompressible condition2}), the linearized 
equation for the conservation of the entropy (\ref{eq: adiabatic condition3}) 
is replaced by
\begin{eqnarray}
\frac{\partial \delta P}{\partial t} + \delta v_{y} \biggl ( \frac{\partial P}{\partial y} 
+ \frac{\Gamma}{\gamma} \frac{\partial \gamma}{\partial y} P \biggr ) = 0 \;. \label{eq: adiabatic condition4}
\end{eqnarray}
This indicates that in the temporal evolution of the pressure
perturbations is triggered by the gradients in the $y-$direction of
both the pressure and also the Lorentz factor. The $y-$variation of
the Lorentz factor is driven by the velocity perturbations $\delta v_y$. 
Since the Lorentz factor is uniform in both the jet and cocoon regions, 
the gradient of the Lorentz factor in the $y-$direction should be considered 
in our model as acting across the jet interface when we derive the 
dispersion relation for the relativistic RTI.

\subsection{Dispersion relation}
\label{sec:dispersion}
We assume that the perturbations of physical variables have the
Wentzel--Kramers--Brillouin (WKB) spatial and temporal dependence
given by ${\rm exp}[i(kx-\omega t)]$, where $k$ and $\omega$ are the
wavenumber in the $x-$direction and the frequency, respectively. Since 
we neglect the impact of the curvature of the jet interface, the wavelength 
of the perturbation in the $x-$direction is much smaller than the jet radius. 
Therefore, using equation~(\ref{eq: pressure scale height2}), we obtain
\begin{eqnarray}
\frac{1}{k}  \ll H \;. \label{eq: WKB1}
\end{eqnarray}
In addition, the evolution of the perturbed system is faster than the
dynamical time of the unperturbed background state,
\begin{eqnarray}
\frac{1}{\omega}  \ll \sqrt{\frac{H}{g}} \;, \label{eq: WKB2}
\end{eqnarray}
where $\sqrt{H/g}$ corresponds roughly to the oscillation time scale
as can be inferred from equations~(\ref{eq: effective gravity}) and 
(\ref{eq: pressure scale height2}). The radial oscillations of the jet 
are advected by the underlying beam in the $z-$direction, with an 
advection speed equal to the unperturbed vertical component in the 
jet velocity, $v_z$. The temporal and spatial scales in the jet direction 
are comparable to or larger than those in the radial direction. Therefore, 
the changes of the background system in the $z-$direction are also 
slow and long compared to those of the perturbed system from 
equations (\ref{eq: WKB1}) and (\ref{eq: WKB2}). A posteriori, this 
justifies neglecting the derivatives of the physical variables with respect 
to $z$.
Considering the WKB ansatz, equations 
(\ref{eq: incompressible condition2})--(\ref{eq: equation of motion2}) 
and (\ref{eq: adiabatic condition4}) become
\begin{eqnarray}
-i \omega \gamma \delta \rho + \delta v_{y} {\rm D}_{y} (\gamma \rho) = 0 \;, \label{eq: incompressible condition3}
\end{eqnarray}
\begin{eqnarray}
i k \delta v_{x} + {\rm D}_{y} \delta v_{y} = 0 \;,
\end{eqnarray}
\begin{eqnarray}
-i \omega \gamma^2 \rho h \delta v_{x} = - i k \delta P \;, \label{eq: equation of motion x}
\end{eqnarray}
\begin{eqnarray}
-i \omega \gamma^2 \rho h \delta v_{y} = -{\rm D}_{y} \delta P 
- \gamma^2 \biggl ( \delta \rho + \frac{\Gamma}{\Gamma -1} \frac{\delta P}{c^2} \biggr ) g \;,
\label{eq: equation of motion3}
\end{eqnarray}
\begin{eqnarray}
-i \omega \delta P + \delta v_{y} \biggl ( - \frac{\Gamma - 1}{\Gamma} \frac{\rho c^2 h}{H} 
+ \Gamma \frac{{\rm D}_{y} \gamma}{\gamma} P \biggr ) = 0 \;, \label{eq: adiabatic condition5}
\end{eqnarray}
where $D_{y} := \partial / \partial y$. Note that the second 
term in equation~(\ref{eq: adiabatic condition5}) is given by using 
equations~(\ref{eq: balancing equation}) and (\ref{eq: pressure scale height1}). 
Combining equations~(\ref{eq: equation of motion3}) and 
(\ref{eq: adiabatic condition5}) leads to
\begin{eqnarray}
- i \omega \gamma^2 \rho h \delta v_{y} = - {\rm D}_{y} \delta P 
- \gamma^2 \delta \rho g + \frac{\gamma^2}{i \omega} \biggl ( \frac{\rho h}{H} 
- \frac{{\rm D}_{y} \gamma}{\gamma} \frac{\Gamma^2}{\Gamma -1} \frac{P}{c^2} \biggr )\, g \, \delta v_{y} \;.
\label{eq: equation of motion4}
\end{eqnarray}
Eliminating the perturbed density $\delta \rho$, pressure 
$\delta P$ and $x$-component of velocity $\delta v_{x}$ in 
equations~(\ref{eq: incompressible condition3})--(\ref{eq: equation of motion x}) 
and (\ref{eq: equation of motion4}), we obtain the following differential 
equation for the $y$-component of velocity in the perturbed state 
$\delta v_{y}$:
\begin{eqnarray}
{\rm D}_{y} (\omega^2 \gamma^2 \rho h {\rm D}_{y} \delta v_{y}) 
- k^2 \omega^2 \biggl ( 1 - \frac{g}{H \omega^2} \biggr ) \gamma^2 \rho h \delta v_{y} \nonumber \\
= k^2 \gamma \biggl [ {\rm D}_{y} (\gamma \rho) + {\rm D}_{y} \gamma \frac{\Gamma^2}{\Gamma - 1} \frac{P}{c^2} \biggr ] \,g\, \delta v_{y} \;.
\end{eqnarray}
Comparing the typical time scales of the perturbed and 
unperturbed system, from equation~(\ref{eq: WKB2}), 
the above differential equation reduces to
\begin{eqnarray}
{\rm D}_{y} (\omega^2 \gamma^2 \rho h {\rm D}_{y} \delta v_{y}) - k^2 \omega^2 \gamma^2 \rho h \delta v_{y} 
\nonumber \\
= k^2 \gamma \biggl [ {\rm D}_{y} (\gamma \rho) + {\rm D}_{y} \gamma \frac{\Gamma^2}{\Gamma - 1} \frac{P}{c^2} \biggr ] \,g\, \delta v_{y} \;.
\label{eq: differential equation}
\end{eqnarray}
Since the density and Lorentz factor are uniform in the jet and cocoon
regions, we may drop the terms $D_y(\gamma\rho)$ and
$D_y\gamma$ on the right hand side of equation~(\ref{eq: differential equation}).
Using equation~(\ref{eq: pressure distribution}), we obtain
\begin{eqnarray}
\rho h = \biggl ( \rho + \frac{\Gamma}{\Gamma - 1} \frac{P_0}{c^2} \biggr ) e^{-y/H} \;,
\end{eqnarray}
relation that can be plugged into equation~(\ref{eq: differential equation}) 
for both regions of the fluid, leading to  
\begin{eqnarray}
{\rm D}_{y}^2 \delta v_{y} - \frac{1}{H} {\rm D}_{y} \delta v_{y} -
  k^2 \delta v_{y} = 0 \;,
\label{eq:dispersionrel}
\end{eqnarray}
where we have assumed that $\omega$ is constant. The general 
solution of equation\,(\ref{eq:dispersionrel}) is
\begin{eqnarray}
\delta v_{y} = Ae^{\alpha_1 y} + Be^{\alpha_2 y} \;, \label{eq: general solution}
\end{eqnarray}
where
\begin{eqnarray}
\alpha_1 = -k \biggl ( -\frac{1}{2kH} + \sqrt{\frac{1}{4k^2H^2} + 1} \biggr ) \label{eq: alpha1}
\end{eqnarray}
and
\begin{eqnarray}
\alpha_2 = k \biggl ( \frac{1}{2kH} + \sqrt{\frac{1}{4k^2H^2} + 1} \biggr ) \;. \label{eq: alpha2}
\end{eqnarray}
From equation~(\ref{eq: WKB1}), we drop $1/2kH$ and 
$1/4k^2H^2$ in equations~(\ref{eq: alpha1}) and 
(\ref{eq: alpha2}). Assuming $\delta v_{y}$ vanishes 
when $y \to \pm \infty$ for the boundaries of $\delta v_{y}$, 
equation~(\ref{eq: general solution}) is replaced by
\begin{eqnarray}
\delta v_{y} = Ae^{-ky} \;\;\; (y > 0) \label{eq: solution1}
\end{eqnarray}
and
\begin{eqnarray}
\delta v_{y} = Ae^{ky} \;\;\; (y < 0) \;. \label{eq: solution2}
\end{eqnarray}
To ensure the continuity of $\delta v_{y}$ across the jet 
interface ($y=0$), the same constant $A$ is chosen in the 
solutions for $y > 0$ and $y < 0$. 

The dispersion relation that we require is obtained by plugging
equations~(\ref{eq: solution1}) and (\ref{eq: solution2}) into
equation~(\ref{eq: differential equation}) and then integrating
equation~(\ref{eq: differential equation}) over an infinitesimal
element of $y$ across the interface and dropping the integral of the
non-divergent term:
\begin{eqnarray}
\omega^2 &=& - gk \frac{\gamma_1^2 \rho_1 h_1 - \gamma_2^2 \rho_2 h_2 + \Gamma (\gamma_1^2 - \gamma_2^2)P_0/c^2}{\gamma_1^2 \rho_1 h_1 + \gamma_2^2 \rho_2 h_2} \label{eq: dispersion relation1} \\
&=& - gk \frac{\gamma_1^2 \rho_1 h_1^{\prime} - \gamma_2^2 \rho_2 h_2^{\prime}}
{\gamma_1^2 \rho_1 h_1 + \gamma_2^2 \rho_2 h_2} \;. \label{eq: dispersion relation2}
\end{eqnarray}
Here, the subscripts $1$ and $2$ stand for the physical variables in the jet and cocoon region, respectively, and 
\begin{eqnarray}
h^{\prime} := 1 + \frac{\Gamma^2}{\Gamma -1} \frac{P_0}{\rho c^2} \;.
\end{eqnarray}
Note that $P_0$ is the pressure at the jet interface. Its coefficient 
in the above expression for $h^{\prime}$ is larger by a factor of 
$\Gamma$ than that in the corresponding expression for 
specific enthalpy $h$ (see equation~\ref{eq: specific enthalpy}). 
This is due to considering the difference in the Lorentz factor 
across the jet interface in equation~(\ref{eq: adiabatic condition4}) 
although there is no difference in pressure. 

Instability sets in when $\omega^2<0$. Thus, from
equation~(\ref{eq: dispersion relation2}), the condition for the 
onset and growth of the RTI at the relativistic jet interface is 
given by
\begin{eqnarray}
\gamma_1^2 \rho_1 h_1^{\prime} > \gamma_2^2 \rho_2 h_2^{\prime} \;. \label{eq: onset condition}
\end{eqnarray}
The important point for this onset condition is that the difference 
in effective inertia between two different fluids does not give 
a criterion for the onset of the RTI. Note that besides the 
difference in the effective inertia, an additional term 
$\Gamma (\gamma_1^2 - \gamma_2^2)P_0/c^2$ is necessary 
for the criterion for the onset of the relativistic RTI.

In the non-relativistic limit ($\gamma \to 1$, $h \to 1$ and $h^{\prime} \to 1$), 
the relativistic dispersion relation for the RTI 
(equation~\ref{eq: dispersion relation2}) corresponds to the 
classical one \citep{Chandrasekhar61}:
\begin{eqnarray}
\omega^2 = - gk \frac{\rho_1 - \rho_2}{\rho_1 + \rho_2} \label{eq: classical dispersion relation1} \;.
\end{eqnarray}
When we consider the non-relativistic flow of relativistically hot gas \citep[$h > 1$ and $\gamma=1$,][]{Allen84}, 
this reduces to 
\begin{eqnarray}
\omega^2 = - gk \frac{\rho_1 - \rho_2}{\rho_1 + \rho_2 +2\Gamma/(\Gamma - 1)P_0/c^2} \label{eq: classical dispersion relation2} \;.
\end{eqnarray}
In both classical cases, the difference in rest-mass density 
between the two fluids drives the RTI.

The temporal growth rate $\sigma$ is defined as the imaginary 
part of the frequency:
\begin{eqnarray}
\sigma := {\rm Im} \; \omega \;. \label{eq: growth rate}
\end{eqnarray}
The dimensionless growth rate is given by
\begin{eqnarray}
\frac{\sigma}{\sqrt{gk}} = \sqrt{\mathcal{A}} \;, \label{eq: dimensionless growth rate}
\end{eqnarray}
where $\mathcal{A}$ is the Atwood number, which is a 
non-dimensional parameter that characterizes the linear 
growth of the RTI. In the relativistic case, from 
equation~(\ref{eq: dispersion relation1}), one can find
\begin{eqnarray}
\mathcal{A} = \frac{\gamma_1^2 \rho_1 h_1 - \gamma_2^2 \rho_2 h_2 + 
\Gamma (\gamma_1^2 - \gamma_2^2)P_0/c^2}{\gamma_1^2 \rho_1 h_1 + \gamma_2^2 \rho_2 h_2} \;.
\label{eq: Atwood number}
\end{eqnarray}
When the Lorentz factor of the jet is much larger than 
that of the cocoon ($\gamma_1 \gg \gamma_2$) and 
the pressure at the jet interface is relativistic 
($P_0 \gg \rho_1 c^2$, $\rho_2 c^2$), the Atwood number 
is almost equal to the ratio of specific heats:
\begin{eqnarray}
\mathcal{A} \sim \Gamma \;.
\end{eqnarray}
This is greater than unity when we consider $\Gamma > 1$, 
for example, the ideal-gas case ($\Gamma = 4/3$). In contrast, 
the classical Atwood number is always less than unity, 
because its denominator is greater than its numerator (see 
equations~\ref{eq: classical dispersion relation1} and 
\ref{eq: classical dispersion relation2}).

\section{Numerical Study of Stability of Interface between Jet and Cocoon}
We perform numerical simulations to investigate the stability 
of the interface separating the jet from the cocoon and the 
growth of the RTI at the jet interface. In particular, we verify 
the dispersion relation of the RTI derived analytically in the 
previous section by comparing the linear growth rates in the 
analytic model and the numerical simulations.

\subsection{Governing equations}
The set up we consider in this section is almost the same as 
that in Section~2.1. Figure~\ref{fig1}c shows the initial geometry 
of the jet--cocoon-medium system schematically. The jet beam 
is on top of the cocoon and it is subject to a uniform acceleration 
driven by the restoring force sketched with a thick black arrow in 
Fig.\,~\ref{fig1}c. We solve the evolution of this system numerically 
by assuming a small amplitude for the corrugated jet interface.

Assuming an ideal gas equation of state with a constant ratio of 
specific heats $\Gamma = 4/3$, the governing equations to be 
solved are
\begin{eqnarray}
\frac{\partial}{\partial t}(\gamma \rho) + \frac{\partial}{\partial x}(\gamma \rho v_{x}) + 
\frac{\partial}{\partial y}(\gamma \rho v_{y}) = 0 \;, \label{eq: mass conservation2}
\end{eqnarray}
\begin{eqnarray}
\frac{\partial}{\partial t}(\gamma^2 \rho h v_{x}) + \frac{\partial}{\partial x}(\gamma^2 \rho h v_{x}v_{x} + P) 
+ \frac{\partial}{\partial y}(\gamma^2 \rho h v_{x}v_{y}) = 0 \;,
\end{eqnarray}
\begin{eqnarray}
\frac{\partial}{\partial t}(\gamma^2 \rho h v_{y}) + \frac{\partial}{\partial x}(\gamma^2 \rho h v_{y}v_{x}) 
+ \frac{\partial}{\partial y}(\gamma^2 \rho h v_{y}v_{y} + P) = -\gamma^2 \rho h g \;, 
\end{eqnarray}
\begin{eqnarray}
\frac{\partial}{\partial t}(\gamma^2 \rho h v_{z}) + \frac{\partial}{\partial x}(\gamma^2 \rho h v_{z}v_{x}) 
+ \frac{\partial}{\partial y}(\gamma^2 \rho h v_{z}v_{y}) = 0 \;, \label{eq: momentum conservation}
\end{eqnarray}
\begin{eqnarray}
\frac{\partial}{\partial t}(\gamma^2 \rho h c^2 - P) + \frac{\partial}{\partial x}(\gamma^2 \rho h c^2 v_{x})
+ \frac{\partial}{\partial y}(\gamma^2 \rho h c^2 v_{y}) = -\gamma^2 \rho h g v_{y} \;, \label{eq: energy conservation}
\end{eqnarray}
where the symbols are defined as in Section~2. Any derivatives 
of a physical variable in the $z$-direction are assumed to be zero. 
The impact on the system of the inertia force in the $y-$direction 
is included in the source terms in both the momentum and energy 
conservation equations. The time evolution of $v_{z}$ and the 
temporal variation of the pressure are considered in 
equations~(\ref{eq: momentum conservation}) and 
~(\ref{eq: energy conservation}), respectively, although we did not 
take them into account in the previous section.

A relativistic HLLC scheme \citep{Mignone05} is used to solve
equations~(\ref{eq: mass conservation2})--(\ref{eq: energy
conservation}) in conserved form. The primitive variables are
calculated from the conservative variables following the method of
\citet{Mignone07}. Second order accuracy is obtained in our code 
by employing an spatial MUSCL-type intercell reconstruction
and a second-order Runge--Kutta time integration. See
\citet{Matsumoto12} and \citet{Matsumoto13} for the details of our
special relativistic hydrodynamic (SRHD) code.

\subsection{Initial setting of jet--cocoon-medium system}
\begin{table}
\begin{center}
\caption{
Rest-mass density of the cocoon in hydrostatic equilibrium, $\rho_2$, for all 
models is listed. The rest-mass density of the jet and the Lorentz factor of the jet 
and the cocoon are fixed in all models. We set $\rho_1=0.1$, $\gamma_1=5$, 
and $\gamma_2=1$. The corresponding relativistic Atwood number ${\cal A}$ for 
each model is also listed. In addition, the dimensionless growth rate 
$\sigma/\sqrt{gk}$ evaluated from numerical results are shown in this table. 
Here, $gk = 2\pi \times 10^{-4}$.
}
\begin{tabular}{lccc}
\hline
 & $\rho_{2}$ & $\mathcal{A}$ & $\sigma/\sqrt{gk}$\\
\hline
Model A (fiducial) & $1$ & $1.2$ & $1.081$\\
Model B & $12$ & $1.0$ & $0.969$\\
Model C & $25.2$ & $0.8$ & $0.892$\\
Model D & $41.6$ & $0.6$ & $0.753$\\
Model E & $62.8$ & $0.4$ & $0.625$\\
Model F & $91.0$ & $0.2$ & $0.422$\\
\label{table1}
\end{tabular}
\end{center}
\end{table}
\begin{figure*}
\begin{center}
\scalebox{0.6}{{\includegraphics{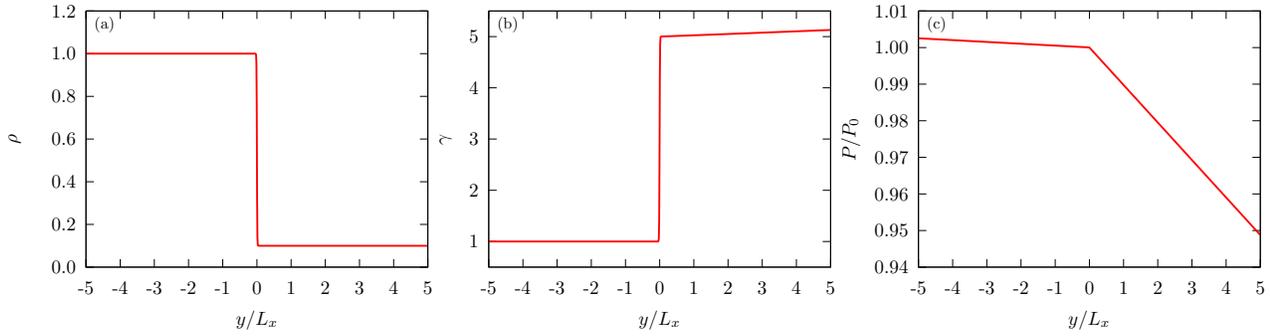}}} 
\caption{Initial spatial distributions of (a) rest-mass density, 
(b) Lorentz factor and (c) pressure in the $y-$direction for the 
fiducial model when $\Delta x / L_x = w / L_x = 0.01$. }
\label{fig2}
\end{center}
\end{figure*}
The jet--cocoon system is initially in hydrostatic equilibrium in the 
$x$--$y$ plane (see Section~2.2 and Fig.~\ref{fig1}c for details). 
The computational domain consists of a rectangle with $x-$~and 
$y-$dimensions $L_x$ and $L_y=10L_x$, respectively, covered 
with a uniform grid with the same mesh spacing in both directions 
of ($\Delta x = \Delta y$). Three different resolutions ($\Delta x/L_x 
= 0.0025$, $0.005$ and $0.01$) are used to test the convergence 
of the growth rate of the RTI for the fiducial model. For other models, 
we use $\Delta x/L_x = 0.01$. Periodic boundary and outflow (zero 
gradient) conditions are set in the $x-$~and $y-$directions, respectively.

The normalization units for length, velocity, time and energy density 
are chosen as $L_x$, the speed of light $c$, the light-crossing time 
over the length of the calculation domain in the $x-$direction, $L_x/c$, 
and the pressure at the jet interface, $P_0$, respectively. In the following,
we set $c=1$.  The upper ($0 < y < 5L_x$) and lower ($-5L_x < y < 0$) 
regions are filled with the jet and cocoon media, respectively. We set 
$P_0=1$ at the jet interface located at $y=0$, and at every other point 
in the domain, we employ equation (\ref{eq: pressure distribution}). 
The jet propagation direction is taken to be the $z-$direction. Following 
the convention stated in the previous section, subscripts $1$ and $2$ 
stand for the physical variables in the jet and cocoon regions, respectively. 
In the following, we fix the pressure at the jet interface, $P_0$, the 
physical variables of the jet ($\gamma_1$ and $\rho_1$) and the Lorentz 
factor of the cocoon, $\gamma_2$. The Lorentz factor and rest-mass 
density of the jet are $\gamma_1 = 5$ and $\rho_1 = 0.1$, respectively. 
This sets the jet as mildly relativistic in terms of internal energy. This 
initialization has been set up for numerical convenience, but our stability 
analysis does not critically depend on the exact ratio of kinetic to 
thermal energy in the jet or in the cocoon, as we shall see.
The Lorentz factor of the cocoon is $\gamma_2=1$. This means that the 
velocity of the cocoon is zero, consistent with the fact that the cocoon is 
in hydrostatic equilibrium. The rest-mass density of the cocoon, $\rho_2$, 
depends on the model. In the fiducial model, we set $\rho_2=1$, which 
makes this model mildly relativistic from the thermodynamics viewpoint.
In the growth of the RTI, the Atwood number $\mathcal{A}$ is an
important parameter with which to investigate the evolution of the system. 
We set $\mathcal{A}=1.2$ in the fiducial model (equation \ref{eq: Atwood number}).
In order to compare the dimensionless growth rates of the analytic 
model and the numerical simulations, we consider six cases: 
$\mathcal{A} = 0.2$, $0.4$, $0.6$, $0.8$, $1.0$ and $1.2$. The Atwood 
number and the corresponding density of the cocoon in hydrostatic 
equilibrium for all models are listed in Table~1. The set of models we 
consider span a useful range of Atwood numbers by changing the value 
of the density of the cocoon ($\rho_2$). For this range of Atwood numbers, 
it turns out that the cocoons of our relativistic jets are only mildly relativistic 
or subrelativistic, since $P_0/\rho_2 \lesssim 1$.

As described in Section~2.2, the inertia force originates from the radial 
oscillations of the jet. From equations~(\ref{eq: pressure scale height1}) 
and (\ref{eq: pressure scale height2}), we can obtain the following
relationship:
\begin{eqnarray}
g L_x \sim \frac{\Gamma}{\Gamma - 1}\frac{1}{\gamma_{\rm  jet}^2}\frac{L_x}{r_{\rm jet}} 
= 10^{-4} \biggl ( \frac{5}{\gamma_1} \biggr )^2 \biggl ( \frac{100}{r_{\rm jet}/L_x}  \biggr ) \;.
\end{eqnarray}
Since the impact of the curvature of the jet interface is neglected 
in this study, assuming $r_{\rm jet} = 100L_x$, we set the normalized 
acceleration $g L_x$ to $10^{-4}$. 

The jet--cocoon system is disturbed by a small amplitude 
perturbations of the $y-$component of velocity in the form of
\begin{eqnarray}
v_y (x,y) = \frac{\delta v}{4} \bigg [ 1 - {\rm cos} \biggl ( \frac{2 \pi x}{L_x} \biggr ) \biggr ] 
\biggl [ 1 + {\rm cos} \biggl ( \frac{2 \pi y}{10L_x} \biggr ) \biggr ]  \;, 
\end{eqnarray}
corresponding to $k=2\pi/L_x$. Indeed, the horizontal domain length, 
$L_x$, is set to coincide with a full wavelength of the perturbation 
introduced. We take $\delta v=10^{-4}$ in all our models. To exclude 
the growth of random perturbations with wavelengths of the order of 
the grid size, we set smooth transition profiles for the density and 
effective inertia in the $y-$direction at the jet interface as follows:
\begin{eqnarray}
\rho (y) = \frac{1}{2} \biggl [ \rho_1 + \rho_2 + ( \rho_1 - \rho_2 ) {\rm tanh}  \biggl ( \frac{y}{w} \biggr ) \biggr ] \;,
\label{eq: density distribution}
\end{eqnarray}
\begin{eqnarray}
I(y) = \frac{1}{2} \biggl [ \gamma_1^2 \rho_1 h_1 + \gamma_2^2 \rho_2 h_2 + ( \gamma_1^2 \rho_1 h_1 - \gamma_2^2 \rho_2 h_2 ) {\rm tanh}  \biggl ( \frac{y}{w} \biggr ) \biggr ] \;. \label{eq: distribution of effective inertia}
\end{eqnarray}
Here, $I(y)$ is the spatial distribution of the effective inertia in
the $y-$direction. $w$ is a parameter that controls the width of 
the jet interface, which should be much smaller than the wavelength 
of the perturbation: we set $w/L_x = 0.01$. The effective width of 
the interface is roughly $4w$ considering the functional dependence 
of the transition layer from equation~(\ref{eq: density distribution}). 
Since the spatial distribution of the effective inertia in the $y-$direction 
is given by equation~(\ref{eq: distribution of effective inertia}), the pressure 
distribution in hydrostatic equilibrium in the calculation domain is obtained 
by integrating the balancing equation (\ref{eq: balancing equation}):
\begin{eqnarray}
P(y) = \int - I(y) g {\rm d}y \;.
\end{eqnarray}
Using the spatial distributions of the rest-mass density, effective 
inertia and pressure, the Lorentz factor also has a smooth transition 
profile as follows:
\begin{eqnarray}
\gamma(y) = \sqrt{\frac{I(y)}{\rho(y) + 4P(y)/c^2}} \;.
\end{eqnarray}
For the fiducial model, the initial spatial distributions of the density,
the Lorentz factor and the pressure in the whole calculation domain are 
shown in Figs.~\ref{fig2}a, ~\ref{fig2}b and ~\ref{fig2}c, respectively. 

\subsection{Results}
\begin{figure*}
\begin{center}
\scalebox{0.43}{{\includegraphics{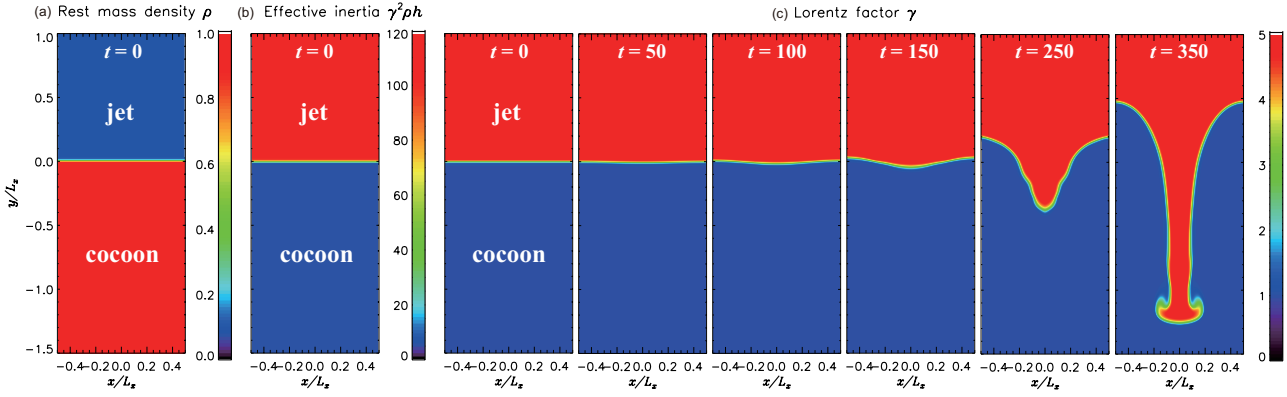}}} 
\caption{Panel (a): Initial spatial distribution of the rest-mass
density. Panel (b): Initial spatial distribution of effective inertia. 
Panel (C): Temporal evolution of the spatial distribution of Lorentz 
factor around the jet interface in the $x$--$y$ plane for the fiducial
model. The relativistic jet is directed in $z$-direction (pointing 
towards the reader from the plane of the page). The jet beam and 
the cocoon are represented in the upper and lower parts of the 
represented panels.}
\label{fig3}
\end{center}
\end{figure*}
The growth of the single RT-mode at the
interface of the relativistic jet is investigated by two-and-a-half
dimensional SRHD simulations. In order to verify the dispersion
relation (\ref{eq: dispersion relation2}) derived in the previous
section, we focus mainly on the linear growth phase of the RTI 
in the numerical simulations, comparing the temporal growth rates 
in the analytical model and in the simulations.

The initial spatial distributions of the rest-mass density and
effective inertia around the jet interface in the $x$--$y$ plane for
the fiducial model with the resolution $\Delta x / L_x = 0.01$ are
shown in panels (a) and (b) of Fig.~\ref{fig3}, respectively. The
inertia force and the relativistic jet are directed in the negative
$y-$direction and the $z-$direction, respectively. The lower-density
medium of the jet is located above the higher-density medium of the 
cocoon. In the non-relativistic regime, the RTI is not expected to grow 
in such a case. However, the effective inertia of the jet is larger than 
that of the cocoon, even though the jet-to-cocoon density ratio 
$\rho_1/\rho_2<1$. Since the jet has a relativistic velocity and is 
relativistically hot whereas the cocoon has a non-relativistic velocity 
and is mildly hot in this model, the Lorentz factor and the relativistic 
thermal energy help to enhance the inertia of the jet. Therefore, the 
effectively heavy medium of the jet is on top of the cocoon medium 
against which the inertia force is pointing. In such a relativistic situation, 
the RTI can grow at the jet interface. Figure~\ref{fig3}c shows the 
temporal evolution of the spatial distribution of the Lorentz factor.  
The amplitude of the corrugated interface grows with time because 
of the growth of the RTI during the linear phase ($t \lesssim 150$). 
We find a mushroom-like structure in the nonlinear regime ($t=350$) 
of the RTI.

\begin{figure}
\begin{center}
\scalebox{0.85}{{\includegraphics{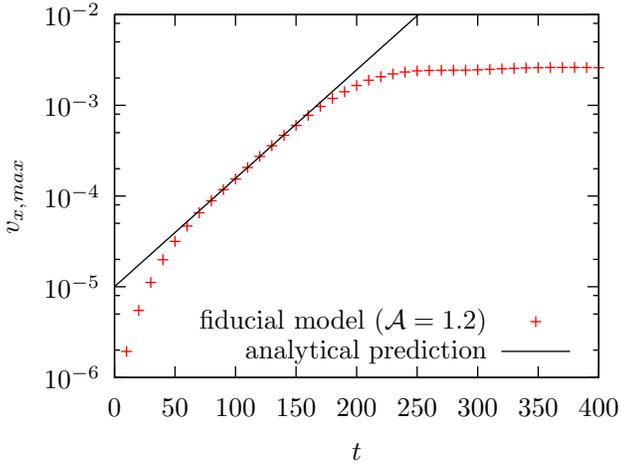}}} 
\caption{Temporal evolution of the maximum $x-$component of 
velocity in the jet region for the fiducial model ($\mathcal{A} = 1.2$). 
The solid line shows the analytical prediction of $v_{\rm x, max} 
\propto {\rm exp}(\sigma t$).}
\label{fig4}
\end{center}
\end{figure}

The linear growth rates of the relativistic RTI in the analytical model 
and the numerical simulation of the fiducial model are compared in 
Fig.~\ref{fig4}. The vertical and horizontal axes represent the maximum 
$x-$component of velocity in the jet region $v_{\rm x, max}$ and the 
time $t$, respectively. To exclude the impact of the smooth transition 
region between the jet and cocoon on the linear growth rate of the RTI, 
we operatively define the jet beam as the region in which the Lorentz 
factor is greater than 90\% of the Lorentz factor of the jet, that is, 
$\gamma \ge 4.5$. The temporal evolution of $v_{\rm x, max}$ in the 
relativistic jet beam is a good indicator for evaluating the linear growth 
rate of the RTI from the results of the numerical simulations. The value 
of $v_{\rm x, max}$ in the relativistic jet beam saturates at the nonlinear 
stage, whereas the RT bubble and finger are accelerated in the 
$y-$direction by the inertia force even in the nonlinear phase, and the 
amplitude of the $y-$component of the perturbed velocity continues to 
grow. The maximum $x-$component of velocity is plotted every $10$ 
time units (crosses in Fig.~\ref{fig4}). The dispersion relation of the 
relativistic RTI (\ref{eq: dispersion relation2}) predicts $v_{x, max} 
\propto {\rm exp}(\sigma t)$, where $\sigma$ is the temporal growth 
rate given by equation~(\ref{eq: growth rate}). This analytical prediction 
is also shown as the solid line in Fig.~\ref{fig4}. 

\begin{figure}
\begin{center}
\scalebox{0.85}{{\includegraphics{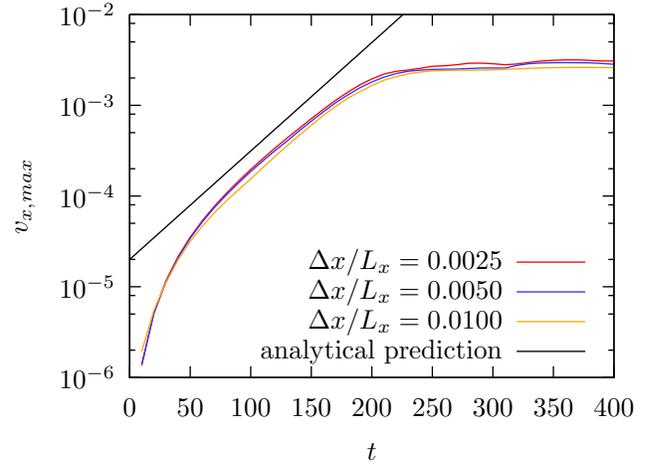}}}
\caption{Grid-resolution dependence on the linear growth rate of 
the RTI for the fiducial model. The initial value of the analytical 
prediction is twice as large as that shown in Fig.~\ref{fig4}.}
\label{fig5}
\end{center}
\end{figure}

In the early phase of the evolution of the disturbed jet interface,
the vertical deformation of the interface occurs between the smooth
transition profiles in the $y-$direction of the density, effective
inertia and Lorentz factor because of the inertia force. This results
in the deceleration of the $y-$component of the perturbed velocity,
contrary to the expectation. The $y-$component of velocity begins to
accelerate after $t \sim 30$. The $x-$component of velocity is
accelerated from the beginning in contrast to the $y-$component of
velocity. However, the eigenstate of the RTI for the single-wavelength
mode we set ($k = 2\pi/ L_x$) is not formed in the deceleration phase
of the $y-$component of velocity. Therefore, the growth rate of $v_{\rm x, max}$ 
is different from the analytical prediction in the early phase of the evolution,
which is a kind of relaxation phase of the initial conditions of the
numerical simulation. After the eigenstate of the single-mode RTI 
is fully achieved ($t \ge 50$), we find in Fig.~\ref{fig4} that the
growth rates in the analytical prediction from the dispersion relation
and in the numerical simulation are almost the same.

The temporal evolution of $v_{\rm x, max}$ in the relativistic jet beam 
for the fiducial model with three different resolutions are shown in 
Fig.~\ref{fig5}. The red, blue and orange solid lines represent the cases 
$\Delta x / L_x = 0.0025$, $0.005$ and $0.01$, respectively. The 
analytically predicted temporal evolution of $v_{\rm x, max}$ is also 
shown by a black solid line, for reference. Note that since we have 
freedom to set the normalization of the relation $v_{\rm x, max} 
\propto \exp{(\sigma t)}$, the initial value of $v_{\rm x, max}$ in the
analytical prediction is taken to be twice as large as that shown in 
Fig.~\ref{fig4}. The growth rate (i.e. the slope of the line during the 
linear regime) converges for the three different resolutions. Therefore, 
the coarsest resolution $\Delta x / L_x = 0.01$ is sufficient for
calculating the linear growth rate of the relativistic RTI for the
fiducial model although the numerical run with the lower resolution
takes slightly longer to achieve saturation of the maximum 
$x-$component of velocity.

\begin{figure}
\begin{center}
\scalebox{0.85}{{\includegraphics{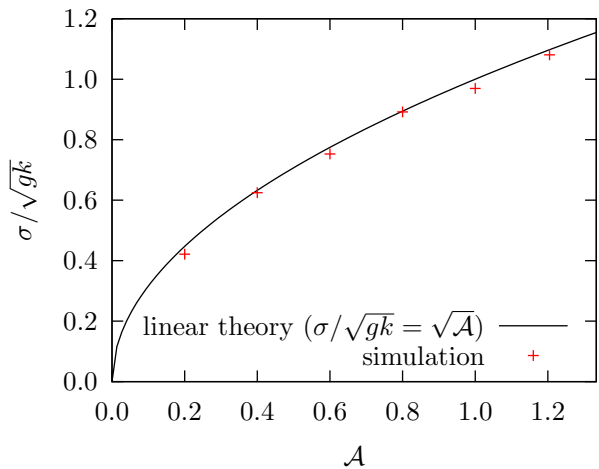}}}
\caption{Comparison of the dimensionless growth rates of the
relativistic RTI in the linear theory and numerical simulations.
The solid line shows the theoretical relation between the
relativistic Atwood number and the dimensionless growth rate 
given by equation~(\ref{eq: dimensionless growth rate}):
$\sigma / \sqrt{gk} = \sqrt{\mathcal{A}}$. Red crosses are results
of numerical runs and their values are listed in Table~1.  }
\label{fig6}
\end{center}
\end{figure}

Figure~\ref{fig6} shows the linear growth rates of the relativistic
RTI in the analytical model and the numerical runs for all models when
$\Delta x / L_x = 0.01$. The vertical and horizontal axes represent
the dimensionless growth rate and the Atwood number, respectively.
The Atwood number characterizes the linear growth of the RTI and is
defined by equation~(\ref{eq: Atwood number}) in the relativistic
jet--cocoon system. The theoretical relationship between the
dimensionless growth rate and the Atwood number is given by
equation~(\ref{eq: dimensionless growth rate}) and is shown by the
solid line in Fig.~\ref{fig6}. The crosses represent the dimensionless
growth rate of the maximum $x-$component of velocity between the
numerical runs; this is evaluated by fitting the data to an exponential 
function. In order to prevent the fits to be affected by the initial numerical 
transient, they are made from a time after which the eigenstate of the
single-mode RTI is fully achieved, as listed in Table~\ref{table1}. 
Note that the Atwood number of the fiducial model, $\mathcal{A}=1.2$, 
is larger than the maximum one in the classical limit (namely, 
$\mathcal{A}=1$; Section~2.4). We find from Fig.~\ref{fig6} that our 
derived dispersion relation predicts the linear growth rate of the 
relativistic RTI correctly, not only for the fiducial model but also for 
the rest models.

\section{Summary and Discussion}
We have studied the linear stability of an interface between the jet 
and the cocoon when the jet is forced to oscillate radially. This radial 
oscillatory motion of the jet is excited naturally by the pressure mismatch 
between the jet and cocoon media as the jet propagates through the 
ambient medium. In the rest frame of the decelerating jet interface that 
is expanding radially, an outward inertia force acts on the interface.  
This situation is physically equivalent to one in which the jet beam is 
supported above the cocoon and the former is subject to an effective 
acceleration. We have performed a linear stability analysis of this system 
in the absence of viscous effects and derived the dispersion relation.

As in the classical case \citep{Chandrasekhar61, Allen84}, the jet 
interface becomes RT unstable when the jet medium is effectively
heavier than the cocoon medium. However, unlike in classical cases,
not only the difference in the rest-mass density between the jet and the 
cocoon, but also the difference in the effective inertia is the relevant 
quantity to drive an RTI in the relativistic regime. Thus, even if the jet is 
lighter than the surrounding medium and pressured matched, a sufficiently 
fast beam becomes effectively heavier than the cocoon and, thus,
it is prone to the RTI.

By using an approximate dispersion relation, \citet{Meliani09} showed 
that a difference in the effective inertia between two fluids separated 
by an interface was a criterion for the onset and growth of the relativistic 
RTI. This is correct when both of fluids are cold, that is, the rest-mass 
energy is much larger than the thermal energy and the pressure does 
not contribute to the fluid inertia. Our derived dispersion relation indicated 
that besides the difference in the effective inertia, an additional pressure 
term was necessary for the general onset condition of the relativistic RTI. 
This additional term originated from the advection of the pressure by the 
gradient of the Lorentz factor across the interface.

The temporal growth rate of the relativistic RTI in the inviscid case 
is also proportional to the square root of the product of the 
wavenumber and effective acceleration as it is in the classical case. 
Therefore, the temporal growth of the shorter-wavelength modes is
faster. On the other hand, the dimensionless growth rate normalized by
$\sqrt{gk}$ is given by the square root of the relativistic Atwood number, 
which is determined by only physical variables of the jet and cocoon. 
The relativistic Atwood number is the most important parameter,
characterizing the linear growth of the relativistic RTI. Its maximum
value is the ratio of specific heats when the Lorentz factor of the
jet is much larger than that of the cocoon and the pressure at the jet
interface is relativistic. The Atwood number is greater than unity in
this case, as opposed to the classical case, for which it is always
smaller than one. This physical condition is expected to be satisfied
when the effective inertia of the jet is much smaller than that of the
ambient medium in which the jet propagates through. This is because in
such a case, the jet is surrounded by a thick and relativistically hot
cocoon heated at the strong reverse shock of the jet head (see
Fig.~\ref{fig1}a).

The relativistic RTI does not grow only at the interface between the 
jet and its cocoon. Once the jet oscillates radially because of the pressure 
mismatch between the jet and a surrounding medium, any jet interface 
except the contact discontinuity at the jet head becomes RT unstable 
\citep[e.g., ``naked jet'',][]{Toma17}. A remarkable result of our study 
is that the interface between different components of the jet is generically 
unstable regardless of the inertia force that drives the RTI. This is due 
to the fact that the instability condition~(\ref{eq: onset condition}) is 
expected to hold in relativistic astrophysical jets. Such inertia force can 
be originated by the pressure gradient considered in this paper or by a 
centrifugal force if the beam flow is rotating \citep[e.g.,][]{Meliani07, 
Meliani09, Marti15, Millas17}. The dispersion relation and growth rate 
of the relativistic RTI derived in this work are also applicable to the 
stability analisys of the surfaces limiting the beam of relativistic flows 
in other astrophysical scenarios \citep[e.g.,][]{Bosch-Ramon15}.

In addition to the linear analysis, the linear growth of the single-mode 
RTI at the interface of the relativistic jet has been investigated by 
two-and-a-half dimensional SRHD simulations within a periodic 
computational box in the direction tangential to the jet interface. 
The numerical set up that we study by means of numerical simulations 
was almost the same as that employed in the linear stability analysis 
of the jet--cocoon system. In our fiducial model, the pressure at the jet
interface is mildly relativistic; the Lorentz factors of the jet and cocoon
are $5$ and $1$, respectively. The corresponding relativistic Atwood
number is $1.2$. The RTI has grown at the jet interface, even though
the beam of the jet has lower rest-mass density than the cocoon in the
fiducial model.  This is because the jet beam is effectively heavier than 
the cocoon because of the enhanced inertia of the former due to its 
larger Lorentz factor and specific enthalpy. The validity of our derived 
dispersion relation is confirmed by a parametric study of different 
cocoons with distinct relativistic Atwood numbers, comparing the linear 
growth rates in the analytical model and numerical simulations.

In this work, we have focused on the linear stability of the jet interface
restricted to motions perpendicular to the jet flow and excluding the 
destabilizing effect of the KHI that grows along the jet direction. 
Further study of the nonlinear regime of the RTI is necessary to 
quantify the stability of the oscillating jet interface. In addition, since 
there shall be a velocity shear at the jet interface in realistic jets, 
the growth of the KHI plays an important role in their stability (see, e.g., 
\citealt{HH03}). The nonlinear evolution of the relativistic RTI at the jet 
interface and the relationship between the KHI and RTI are without 
the scope of our work and will be reported in our subsequent paper. 
However, here we may anticipate an interesting result. We note that 
the instability condition~(\ref{eq: onset condition}) can be rewritten as
\begin{eqnarray}
\eta^*_{\rm Rc}:=\frac{\rho_1 h_1^{\prime} \gamma_1^2}{\rho_2 h_2^{\prime} \gamma_2^2} > 1 
\end{eqnarray}
A parameter formally similar to $\eta^*_{Rc}$ was defined in \cite{Marti97} as the key to differentiate 
morpho-dynamical properties of relativistic jets, namely
\begin{eqnarray}
\eta^*_{\rm R}:=\frac{\rho_1 h_1 \gamma_1^2}{\rho_{\rm a} h_{\rm a} \gamma_{\rm a}^2} \;,
\end{eqnarray}
where the subscript ``a'' refers to quantities of the ambient medium. 
Large values of $\eta^*_{\rm R}$ yield relatively smooth and
featureless jets, while relativistic jets inflate large cocoons and 
develop beams with numerous recollimation shocks in the regime
$\eta^*_{\rm R}\ll 1$, according to \cite{Marti97}. Indeed, \cite{HH03}, 
find that $\eta^*_{\rm R}\gg 1$ is essential to prevent the development 
of the KHI. We note that our parameter $\eta^*_{\rm Rc}$  is linked to
$\eta^*_{\rm R}$ through
\begin{eqnarray}
\eta^*_{\rm Rc} =\eta^*_{\rm R}\frac{h_1^{\prime}}{h_1}
\frac{\rho_{\rm a} h_{\rm a} \gamma_{\rm a}^2}{\rho_2 h_2^{\prime} \gamma_2^2} \;.
\label{eq:etaRc-etaR}
\end{eqnarray}
Here, $h_1^{\prime}/h_1 \simeq 1$. Since typically the ambient 
medium is at rest ($\gamma_{\rm a}=1$), is cold ($h_{\rm a}\simeq 1)$ 
and (much) denser than the cocoon ($\rho_{\rm a}/\rho_2\gg 1$), the 
fraction $\rho_{\rm a} h_{\rm a} \gamma_{\rm a}^2 / (\rho_2 h_2^{\prime}
\gamma_2^2) \simeq \rho_{\rm a}/(\rho_2 h_2^{\prime}\gamma_2^2)$ 
in equation~(\ref{eq:etaRc-etaR}) is typically of the order of or
larger than unity. This means that the regime in which $\eta^*_{\rm
  R}\gg 1$, we also expect $\eta^*_{\rm Rc}\gg 1$. Hence, the regime
in which the KHI is absent, because the effective inertia of the jet
beam is much larger than that of the external medium, is optimal for
the development of the RTI at the jet/cocoon interface although
a driving force for a radial motion of the jet is necessary.

Another interesting consequence of our analysis comes from the fact
that the parameter $\eta^*_{\rm R}=1$ sets the boundary between
non-relativistic and relativistic jet propagation regimes (e.g.,
\citealt{Matzner03,Bromberg11}). Even more,
$\eta^*_{\rm R} >\theta_{\rm j}^{-4/3}\gg 1$ (where $\theta_{\rm j}$
is the jet opening angle), defines the border between uncollimated and
collimated jets (e.g., \citealt{Bromberg11}). With the same
reasoning than in the previous paragraph, in the relativistic jet
propagation regime as well as in the uncollimated jet regime, the
jet/cocoon interface is expected to be RTI.

Finally, we point out that the development of the RTI can be affected
by the existence of gradients in the properties of jets in the across
the jet section. The interaction of extragalactic jets with their
environment leads to the stratification of the beam of the jet in the
direction normal to its velocity \citep[e.g.,][]{PL07, AM08, Hervet17}
In such cases, we do not have a single Atwood number characterizing 
the jet's beam. Instead, a spectrum of Atwood numbers may exist, 
depending on the exact stratification of the hydrodynamic properties 
in the boundary layer between the jet and the external medium. 
We will address the development of the RTI in these cases in a
future work.
\section*{Acknowledgments}
We thank Y.\, Masada, H.\, R.\, Takahashi, A.\, Hillier, A.\, Mizuta, S.\, Nagataki, 
S.\, S.\, Komissarov, A.\, MacFadyen and J.\, M.\, Mart\'{\i} for useful
discussions. The numerical computations were carried out on a Cray
XC30 at the Center for Computational Astrophysics at the National
Astronomical Observatory of Japan and on a Cray XC40 at YITP at Kyoto
University. This work was supported by JSPS KAKENHI Grant Number 
JP 17K14308 and in part by the Center for the Promotion of Integrated 
Sciences (CPIS) of Sokendai. MP acknowledges support by the Spanish 
``Ministerio de Econom\'{\i}a y Competitividad'' grant AYA2013-48226-C3-2-P. 
MAA acknowledges support from the European Research Council 
(grant CAMAP-259276) and the grants AYA2015-66899-C2-1-P and 
PROMETEOII/2014-069.

\label{lastpage}

\end{document}